\documentclass[%
 aip,
 sd,%
 amsmath,amssymb,
 reprint,%
]{revtex4-1}

\usepackage{graphicx}% Include figure files
\usepackage{dcolumn}% Align table columns on decimal point
\usepackage{bm}% bold math
\begin{document}

\preprint{AIP/123-QED}

\title[Bose $et ~al.$]{Fast carrier dynamics in GaAs photonic crystals near the material band edge at room temperature}% Force line breaks with \\

\author{Ranojoy Bose}
\email[]{ranojoy.bose@hp.com}
%\thanks{}
\author{Jason S. Pelc}
\author{Charles M. Santori}
\author{Sonny Vo}
\author{Raymond G. Beausoleil}
\affiliation{Hewlett-Packard Laboratories, Palo Alto, CA, 94304}

\date{\today}% It is always \today, today,

\begin{abstract}
We measure fast carrier decay rates (6 ps) in GaAs photonic crystal cavities with resonances near the GaAs bandgap energy at room temperature using a pump-probe measurement. Carriers generated via photoexcitation using an above-band femtosecond pulse cause a substantial blue-shift in the cavity peak. The experimental results are compared to theoretical models based on free carrier effects near the GaAs band edge. The probe transmission is modified for an estimated above-band pump energy of 4.2 fJ absorbed in the GaAs slab.
%
%Valid PACS numbers may be entered using the \verb+\pacs{#1}+ command.
\end{abstract}

%\pacs{42.65.-k,42.65.Sf}% PACS, the Physics and Astronomy
%                             % Classification Scheme.
%\keywords{GaAs, nonlinear optics, photonic crystals, optical switching}%Use showkeys class option if keyword
%                              %display desired
\maketitle

Nonlinear optical elements are considered to be key components in future optical communication networks \cite{moss_new_2013}. Such networks will require fast, low-power nonlinear optical devices on a robust, scalable platform, and may eventually enable data processing entirely in the optical domain. Optical switching has been demonstrated at attojoule levels using single atomic systems as the nonlinear medium for applications in both classical and quantum information processing, but these systems are usually not viable for large-scale processing \cite{englund_ultrafast_2012, bose_low-photon-number_2012}. A more scalable approach is to use bulk nonlinearities involving free carriers, but enhanced substantially by using photonic cavities that support strong field intensities in small volumes \cite{husko_ultrafast_2009,nozaki_sub-femtojoule_2010,nozaki_ultralow-power_2012,pelc_picosecond_2014,kuramochi_large-scale_2014,lin_all-optical_2014}. These devices offer the benefit of small footprints and may allow large scale integration for applications in optical communication, memory and logic. III-V materials are particularly interesting for many of these applications because sources, detectors and modulators can be engineered on the same material platform. The demonstration of a III-V material all-optical switch that nominally operates below femtojoule energy levels and achieves a high operational speed of 40 Gbps due to fast free carrier recombination has been particularly promising in this respect \cite{nozaki_sub-femtojoule_2010}.

Studying the dynamics of nonlinear processes in single photonic cavities is the first step in the development of advanced photonic circuits. In this study, we investigate free-carrier dynamics in a gallium arsenide (GaAs) photonic crystal cavity that confines light in a small volume ($V \approx (\lambda/n)^3$, where $\lambda$ is the wavelength and $n$ is the refractive index of the medium), and supports a high quality factor optical mode ($Q = 3600$). GaAs cavities are chosen to operate near the direct bandgap energy at 1.42 eV (cavity resonance $\lambda_0 >$ 870 nm). At photon energies near the band gap, one expects enhanced optical response per injected carrier (dn/dN) than at energies well below the band gap \cite{bennett_carrier-induced_1990}. In contrast to previous experiments that use a single laser excitation and quantum dot photoluminescence to observe free carrier dynamics \cite{fushman_ultrafast_2007}, all experiments here are performed without the use of any active materials so that free carrier effects may be unambiguously observed. All experiments are performed at room temperature.

%These effects may combine to reduce the number of photons needed for each operation compared to devices that operate at higher wavelengths and rely only on the plasma effect. In the experiments here, we optically inject carriers above the GaAs band gap energy, and monitor the temporal response of the cavity using a second laser that is swept over the cavity resonance.

The devices used in this work are fabricated on a GaAs wafer that is epitaxially grown by IQE. Two-dimensional photonic crystals are designed for room-temperature operation in the 890-910 nm wavelength range. Photonic crystals are defined using electron beam lithography followed by dry etching. The final devices are formed by undercutting a sacrificial layer of aluminum gallium arsenide (Al$_{0.8}$Ga$_{0.2}$As) underneath the 110-nm GaAs layer, forming a free-standing membrane. A linear defect (three missing air holes) is located at the center of the photonic crystal as shown in Fig. 1a  \cite{akahane_fine-tuned_2005}, allowing for a confined, high-quality resonator mode. The device parameters are as follows: lattice period $a$ = 240 nm, radius $r$ = 70 nm and thickness $t$ = 110 nm.

\begin{figure}[ht]
%\begin{center}

\includegraphics[width=9 cm]{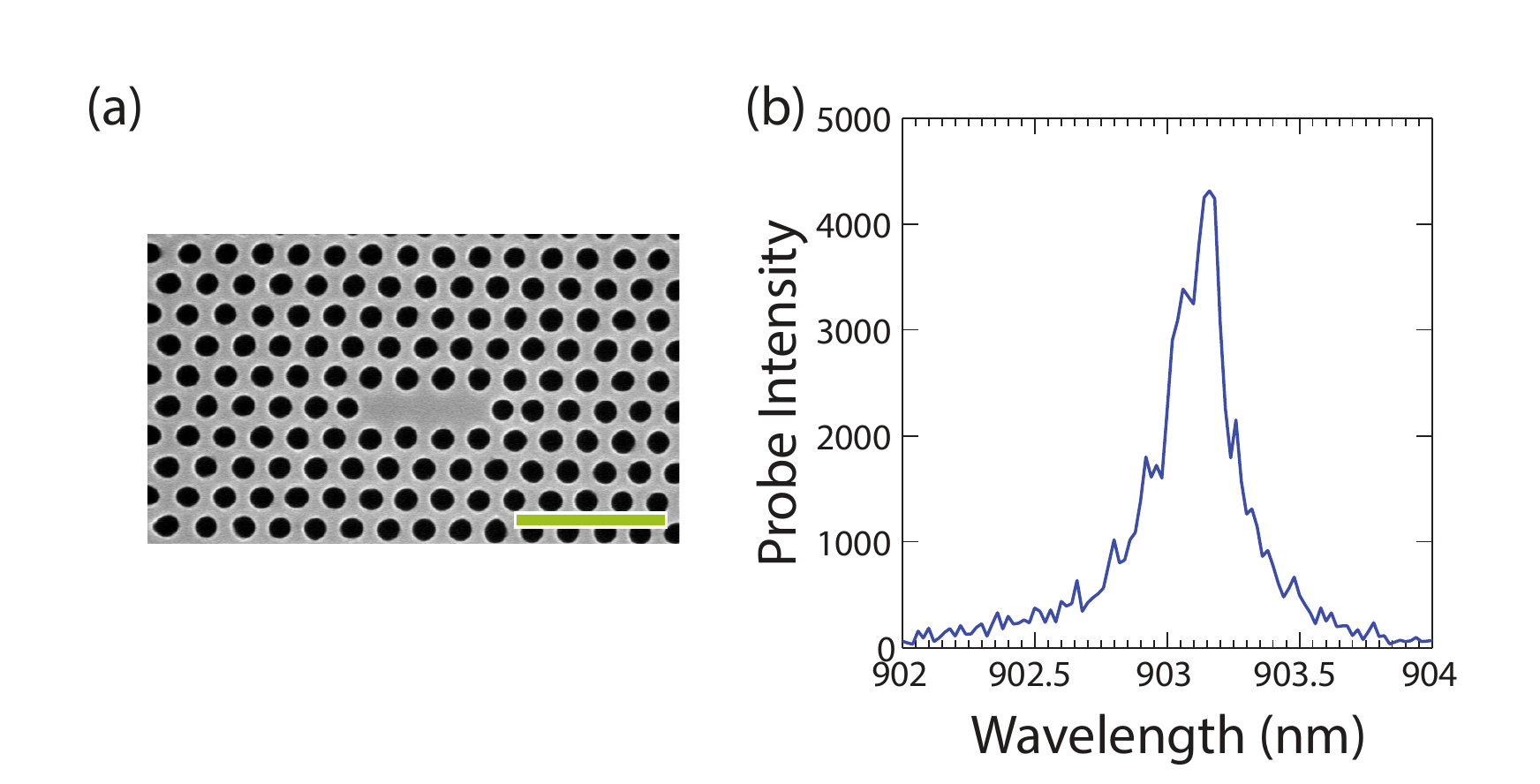}
%\end{center}
\caption{\label{fig:fig1}(a) Scanning electron micrograph of fabricated photonic crystal device. A linear defect cavity is formed by removing 3 holes along a row of the photonic crystal lattice. Scale bar 1 $\mu$m. (b) Low power reflectivity spectrum of cavity optical mode.}
\end{figure}

Measurements are performed at room temperature and atmospheric pressure. The devices are first characterized by their low-power spectra using cross-polarized detection. A low-power 6 $\mu$W tunable laser (New Focus Velocity) is focused vertically on the sample using a 63$\times$ objective lens (1.4 $\mu$m$^2$ laser spot size), and the reflected laser signal is recorded using a spectrometer. In this scheme, the cavity spectrum appears as a peak in the reflected laser spectrum. The cavity spectrum is shown in fig. 1b with a resonance wavelength of 903.2 nm (1.37 eV) and a fitted Q of 3600 (cavity decay rate $\gamma/(2\pi)$=92.5 GHz). The cavity lifetime is given by $\tau = 1/\gamma = $ 1.7 ps.

In order to study nonlinear dynamics in the photonic cavity, carriers are generated in the cavity region using a Ti:saph laser at a pump wavelength  of 796 nm, with a pulse duration of approximately 200 fs and a repetition rate of 75 MHz. The average above-band pump power is 16 $\mu$W after the objective lens. The pump laser is synchronized to a spectrally resolved Hamamatsu streak camera system, with a measured time resolution of 3.4 ps. The photo-generated carriers cause a change in the refractive index in the cavity region, resulting in a blue-shift of the cavity. This shift can be monitored by scanning the continuous wave tunable laser at low powers (probe laser) through the spectral region of the cavity. Individual time-resolved spectra corresponding to different wavelengths of the tunable laser are shown in fig. 2a. The data show that when the probe laser is blue-shifted from the cavity resonance, the arrival of the pump laser results in a peak due to the shift of the cavity that brings it on resonance with the probe laser. On the other hand, when the probe laser is red-shifted from the cavity resonance, the reflected signal shows a pronounced dip due to the cavity blue-shift. In fig. 2b, we show a heat plot where the reflected tunable laser intensity (incident power 6 $\mu$W) is shown as a function of tunable laser wavelength and delay between pump and probe. The probe laser is scanned between wavelengths of 902 nm and 904 nm in increments of 0.02 nm, and at each laser wavelength, the laser signal is integrated for two seconds on the streak camera. The pump-probe measurement simultaneously offers high spectral and temporal resolution. The pump laser arrival time at 22 ps can be directly observed on the streak camera at the pump laser wavelength. The figure shows that after the pump arrival (Delay $\Delta$t $>$ 22 ps), the cavity is considerably blue-shifted, and recovers over a timespan of several tens of ps.

\begin{figure}[ht]
\begin{center}
\includegraphics[width=9 cm]{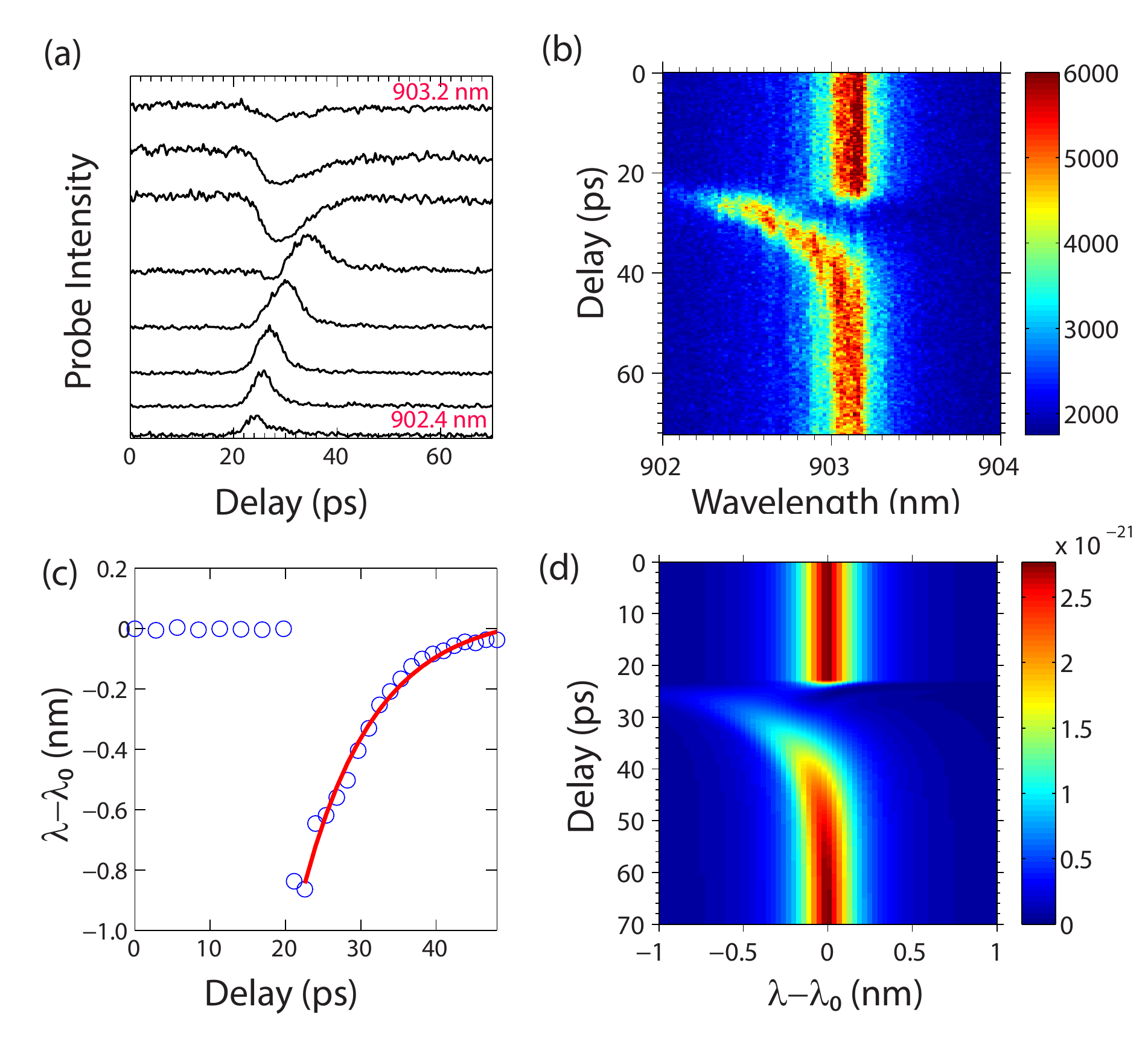}
\end{center}
\caption{\label{fig:fig2}(a) Reflected probe intensity as a function of pump-probe delay for probe laser wavelengths between 902.4 nm and 903.2 nm. (b) Heat plot showing the reflected laser intensity as a function of probe laser wavelength and pump probe delay. The pump laser arrives at 22 ps (c) Cavity maximum as a function of pump-probe delay. Blue circles: Experiment. Red Curve: Exponential fit (d) Simulated cavity dynamics based on experimentally measured parameters.}
\end{figure}

In fig. 2c, we plot the cavity maximum as a function of the delay between pump and probe, by fitting the instantaneous cavity spectra using Lorentzians. The plot shows a total shift of 0.75 nm for the cavity resonance, corresponding to $\Delta \lambda = 3\Delta \lambda_0$, and a 13 dB on/off ratio. Fitting the cavity decay rate to a single exponential gives a recovery time of 6 ps.

We can estimate the photogenerated carrier concentration $N_{exp}$ in our device by using the simple formula
\begin{equation}
%N_\mathrm{ph} = \frac{\lambda^4 \tilde{V}}{2\pi\, n^2 \tau_c\, c\, \xi\, \Gamma\, \eta_c}
%N_{exp}=\frac{P(1-\eta)(1-e^{-\alpha L})\nu}{R_L \hbar \omega_{pump} A L}
N_{\mathrm{exp}}=\frac{P\eta}{R_{\mathrm{L}} \hbar \omega_{\mathrm{pump}} A_{\mathrm{eff}} L}
\end{equation}
where $P$ is the pump laser average power measured after the objective lens, $\eta = 0.14$ is the fraction of incident light that is absorbed in the GaAs photonic crystal slab computed using finite difference time domain simulations, $R_{\mathrm{L}}$  is the repetition rate of the laser, $\hbar$ is the reduced Planck’s constant, $\omega_\mathrm{pump}$ is the pump frequency, A$_{\mathrm{eff}}$ =  1.15 $~\mathrm{\mu m^2}$ is the effective mode area, computed using a Gaussian laser spot profile and accounting for the fact that carriers are not generated in the hole regions of the photonic crystal, and L = 110 nm is the thickness of the GaAs slab. Using the above equation we estimate $N_{\mathrm{exp}} = 1.1 \times 10^{18} $~cm$^{-3}$. From the carrier concentration, we can derive an experimental value of $dn/dN = - (\Delta \lambda)\times n_{\mathrm{g}}/(\lambda N_{\mathrm{exp}} ) = 2.86 \times 10^{-21}$~\text cm$^{3}$, where $n_{\mathrm{g}} = 3.68$ is the group index of light in the GaAs photonic crystal.

The observed dynamics can be simulated by using dynamic coupled mode equations for the cavity field amplitude $a(t)$ and carrier density $N(t)$, given by
\begin{equation}
%%\label{eqn:cmt}
%%\begin{align}
\frac{da}{dt} = \left[ i(\omega_0 - \omega) - i\left( \frac{\omega_0}{n_g} \frac{dn}{dN} \right)N - \frac{\gamma}{2} \right]a + \kappa s(t)\\
%
%%\end{align}
\end{equation}
\begin{equation}
\frac{dN}{dt} = -\frac{N}{\tau_c} + J(t)
\end{equation}
where $\kappa$ is the coupling rate between the external optical input $s(t)$ and the cavity mode $a$, $\tau_{c}$ is the free-carrier lifetime, and $J(t)$ describes the optical carrier injection (200 fs Gaussian pulse), $\omega_0$ is the cavity resonant frequency and $\omega$ is the probe laser frequency.  The total cavity loss rate $\gamma$ incorporates scattering, absorption, and can incorporate free-carrier absorption via a carrier dependence.  We assume a linear change in refractive index with carrier density  $(dn/dN)$ which is shown to be valid for carrier densities below $10^{20}$~cm$^{-3}$ ~\cite{bennett_carrier-induced_1990}.  These simplified equations neglect carrier generation via probe absorption, which is not expected to be a significant.

Fig. 2d plots simulated cavity spectra as a function of pump-probe delay and probe wavelength, based on the experimentally derived nonlinear parameters and  carrier relaxation time. A comparison between the simulations and experimental data show that the system is described well by the coupled mode equations presented above.

The experimental observations are compared to theoretically predicted models for free carrier dynamics near the GaAs bandgap energy. In experiments, we observe only blue shifts in the cavity resonance regardless of pump power. Because of this, we only include bandfilling dispersion (BFD) and plasma dispersion effects in our theoretical model, and ignore bandgap renormalization because it predicts a red-shifting nonlinearity for some pump powers. The theoretical treatment follows ref. [\cite{bennett_carrier-induced_1990}] using material parameters for GaAs under the same physical assumptions. In the case of bandfilling dispersion, the change in refractive index is computed from the theoretical change in the absorption coefficient using Kramers Kronig formalism. In the case of plasma dispersion, the Drude model is used. Following this model, we compute theoretical values of $N = 2.2 \times 10^{17} $~cm$^{-3}$ and $dn/dN = 1.4 \times 10^{-20}$~\text cm$^{3}$ at the probe laser wavelength ($\sim$0.96$E_g$). %A comparison of the two models with the experimental results is shown in Table 1, showing a reasonable agreement between the theory and experimental data for both models.

We note that the theory predicts free-carrier concentrations that are lower than experimental observations by a factor of 5.  The origin of this discrepancy cannot be known for certain based on our experiments, but one possible explanation is that when carriers are pumped above the GaAs bandgap, some carriers relax to the lowest available conduction and valence band states via rapid thermalization, causing band-edge nonlinearities, while other carriers recombine at surfaces or via other trap states and are lost. We did not observe any difference in the measured carrier lifetime by changing the laser spot area between 1 and 20 $~\mathrm{\mu m^2}$, and conclude that the measured lifetime is not influenced by the rate of carrier diffusion in the GaAs photonic crystal.
%The measured cavity carrier lifetime of 6 ps can either be due to the cavity photon decay rate, or due to carrier relaxation from the lowest energy bands, and is likely to be a convolution of the two effects. Experimentally, the data was best fit to a single exponential so that different dynamical effects could not be disambiguated.%

%
\begin{table}
\caption{\label{tab:table1}Comparison of free-carrier density $N$ [\text cm$^{-3}]$  and $dn/dN$ [\text cm$^{3}]$ from theory and experiments  }
\begin{ruledtabular}
\begin{tabular}{ccd}
Effect&$N$&\mbox{$dn/dN$}\\
\hline
%FCA&1.77$\times 10^{18}$&\mbox{1.84$\times 10^{-21}$}\\
Plasma$+$BFD&2.2$\times 10^{17}$&\mbox{1.4$\times 10^{-20}$}\\
%FCA$+$BFD$+$BGR&7.5$\times 10^{17}$&\mbox{1.12$\times 10^{-20}$}\\
\hline
Experiment&1.1$\times 10^{18}$&\mbox{2.86$\times 10^{-21}$}\\
\end{tabular}
\end{ruledtabular}
\end{table}

%We find however, that when all three effects are included, the theoretical model predicts a red-shift in the cavity resonance for carrier concentrations of $ 8 \times 10^{16} $~cm$^{-3}$ $<$ $N < 4 \times 10^{17} $~cm$^{-3}$ caused by the theoretically predicted bandgap renormalization effect near the band-gap energy. However, we were not able to unambiguously reproduce this effect in experiments by reducing the power of our pump laser, and observed a monotonically increasing blue-shift for increasing above-band pump powers. These findings indicate that our experimental findings may be predominantly caused by free carrier absorption and bandfilling dispersion. Using the coupled mode equations, we were able to reproduce the experimental results with the values of $N$ and $dn/dN$ from Table 1 for bandfilling dispersion and free carrier absorption. The discrepancy between the experimental estimate and the theoretically predicted value of $N$ is within our estimate of experimental uncertainties caused by the pump beam focus, carrier losses during thermalization, and inaccuracies in the theoretical models of band filling dispersion and bandgap renormalization.

%DATA FOR THE FOLLOWING FIGURE IS IN A FOLDER CALLED TEST5
% Neff=4.5e17 dndN =1.4e-20 Aeff=5.4 um^2

\begin{figure}
\begin{center}
\includegraphics[width=9 cm]{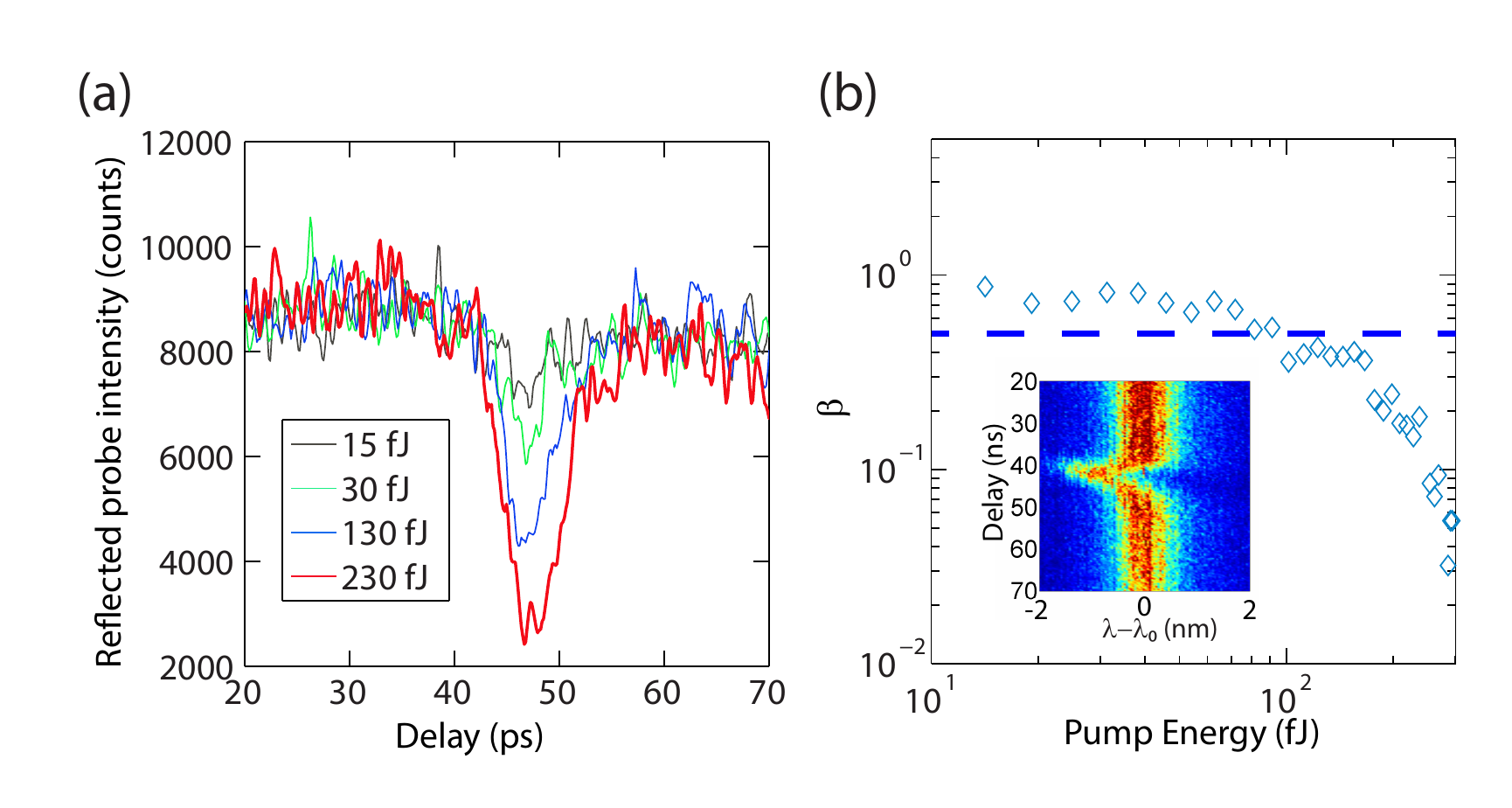}
\end{center}
\caption{\label{fig:fig3}(a) Reflected probe intensity as a function of pump pulse energy when the probe is resonant with the original (low power) cavity mode resonance wavelength. (b) Normalized transmission $\beta$ as a function of pump pulse energy.}
\end{figure}

We finally present experimental results for different powers of the pump laser. Fig. 3a shows temporal scans of the tunable probe laser when fixed at the cavity resonance wavelength with increasing powers of the above-band pump laser. At a low excitation power, the probe laser is mostly reflected, however with increasing pump pulse energy between 15 and 230 fJ measured after the objective lens, the probe laser shows a pronounced dip when the pump laser arrives due to a blue-shift of the cavity resonance. These experiments are performed on a separate but similar device $(Q = 1000, \lambda_0 = 902.9 ~\mathrm{nm}, \Delta \lambda = 1.6 ~\mathrm{nm})$. A heat plot of the probe laser reflectivity with pump-probe delay at 400 fJ incident pump pulse energy is shown in the inset to figure 3b.  We define a normalized transmission parameter $\beta$ based on the maximum and minimum values of the reflected probe laser signal as
\begin{equation}
β\beta(E_{\mathrm{p}}) = \frac{R_{\mathrm{max}}-R(E_{\mathrm{p}})}{R_{\mathrm{max}}-R_{\mathrm{min}}}
\end{equation}
where $R_{\mathrm{max}}$ and $R_{\mathrm{min}}$ are the reflected laser signal at the original cavity resonance for incident pump pulse energies of 300 fJ and 1.6 fJ, respectively, and R is the reflected laser signal at pulse energy $E_{\mathrm{p}}$. In fig. 3b, we plot $\beta$ as a function of the pump pulse energy. Based on these values, we estimate that a value of $\beta = 0.3$ is achieved for incident pulse energy of 30 fJ $($$2 \times 10^4$ absorbed photons, corresponding to 4.2 fJ absorbed pump energy$)$, and $\beta$ = 0.1 is achieved for incident pulse energy of 100 fJ.

In conclusion, we have demonstrated low-power nonlinear carrier dynamics in a nanoscale GaAs photonic crystal cavity. We measure a blue-shifting nonlinear effect with a measured lifetime of 6 ps. We observe a shift in the cavity resonance of 3 times the cavity linewidth by using 213 fJ of incident pump energy. This corresponds to a photo-injected carrier concentration of $1 \times 10^{18}  $~cm$^{-3}$. The probe transmission is modified for ultra-low pump energy of 4.2 fJ, when low absorption in the GaAs slab is taken into account. Our experiments are consistent with free-carrier models based on bandfilling dispersion and the plasma dispersion effect near the material band-edge, when carrier losses are included. In separate measurements, we have achieved absorption-limited Qs exceeding 10$^{4}$ at the wavelength of interest. The observed nonlinear dynamics can be used to demonstrate a low power optical switch by using a below-band pump laser at or near the cavity resonance. Eventually, many of these single nonlinear devices may be connected via integrated optics in advanced optical circuits \cite{santori_quantum_2014}.

\begin{acknowledgments}
This work is supported by the Defense Advanced Research Projects Agency under Agreement No. N66001-12-2-4007. The authors would like to acknowledge Kelley Rivoire and Victor Acosta for helpful discussions. Electron beam lithography was performed at the Stanford Nanofabrication Facilities.
\end{acknowledgments}

%\nocite{*}
\bibliography{Bibtexlibrary10_1_14}% Produces the bibliography via BibTeX.

\end{document}